\documentclass[final]{aipproc}
\layoutstyle{6x9}
\def\laeq{\raise.2ex\hbox{$<$}\kern-.75em\lower.9ex\hbox{$\sim$}\,}
\def\gaeq{\raise.2ex\hbox{$>$}\kern-.75em\lower.9ex\hbox{$\sim$}\,}

\begin{document}

\title{HESS Observations of Pulsar Wind Nebulae}
\classification{97.60.Gb}
\keywords      {pulsar wind nebulae: general -- stars: neutron -- gamma-rays}

\author{Ocker C.\ de Jager}{
  address={Unit for Space Physics, North-West University, Potchefstroom Campus, Private Bag X6001, Potchefstroom, 2520, South Africa}
}

\begin{abstract}
The high resolution capabilities of the {\it High Energy Stereoscopic System} ({\it HESS}) introduced a new era in Gamma-Ray
Astronomy, and opens a new window on pulsar wind nebula (PWN) research. A rotationally induced jet (associated with PSR\,B1509-58) is resolved for the first time in $\gamma$-rays, allowing us to trace the particle transport directly,
without having the complicating effect of spatially varying field distributions on the synchrotron emissivity. 
For PWN older or more extended than Crab (i.e. those with lower field strengths), {\it HESS} also reveal the properties of electrons 
contributing to the EUV/soft X-ray synchrotron bands, whereas EUV/soft X-rays suffer from severe interstellar absorption effects. Finally, {\it HESS} morphological studies of evovled PWN also allow us to directly measure the effects of assymetric
reverse shock interactions due to SNR forward shock expansion into the inhomogeneous interstellar medium.

\end{abstract}

\maketitle

\section{Introduction}
Weiler \& Panagia \citep{WP78} first coined the term "filled-center" or "plerionic" nebulae, following the discovery of
galactic sources with the following features: (1) filled center or blob-like form; (2) a flat radio spectral index
between 0 and -0.3; (3) a well organised magnetic field; and (4) a high integrated linear polarisation at high radio
frequencies. In 1980, Weiler \& Panagia \citep{WP80} added Vela X, part of the Vela SNR, as a member of this class of sources.
The offset of the Vela X PWN relative to the Vela pulsar was puzzling, until resolved as discussed below.

X-ray Astronomy was quite instrumental in detecting and characterising these pulsar wind nebulae (PWN) and we were left with
Radio and X-ray Astronomy contributing mostly to our knowledge of PWN, other than the Crab Nebula.
A few useful contributions from Infrared Astronomy were also made, filling the spectral gap between radio and X-rays.
The radiation mechanism for radio through X-rays is generally believed to be synchrotron emission as a result of pulsar 
injected electrons, with X-rays resulting from freshly injected electrons, 
whereas relic electrons contribute mostly to the radio part. The ordered magnetic field is usually toroidal (due to the spinning
pulsar), with magnetization resulting from the deceleration of the post-shock pulsar wind flow \citep{KC84}. The
synchrotron brightness is a result of a convolution of the electron density and the
magnetic field strength, whereas the same electrons also scatter ambient photons into the high energy
to ultra high energy $\gamma$-ray range. Both components should be observed to understand the dynamics of the pulsar wind.

The well-known optical "wisps" in the Crab Nebula are believed to mark the pulsar wind shock radius, where the pressure
from the unshocked (upstream) pulsar wind balances the (downstream) nebular pressure \citep{KC84}. The high resolution
X-ray images of {\it Chandra} contributed significantly to this field, showing that the presence of a torus and perpendicular
jet are common to PWN (see \citep{NR04} and references therein), with an underluminous region on the inner
part of the torus, marking the PWN shock radius. Ng \& Romani \citep{NR04} managed to derive the geometry
of these structures, which establishes the orientation of the pulsar spin axis (aligned with the
X-ray jet) on the plane of the sky. \emph{Thus, unified pulsar-PWN models
should combine measured radio pulse polarisation swing data and high resolution X-ray PWN geometry for internal consistency.}

A complicating factor for evolved PWN is the effect of the reverse shock on the PWN for SNR expansion
into an inhomogeneous ISM: van der Swaluw et al. \citep{vds01} showed that after the passage of the reverse shock into the PWN, the latter will expand subsonically, settling to a radius, which is $\sim 25$\% of the SNR forward shock radius. Blondin et al. \citep{b01} also showed that an inhomogeneous ISM pressure will result in the reverse shock returning to the PWN at different times of its evolutionary history, 
which can shift the position of the PWN, as observed for Vela X and G\,18.0-0.7 (the latter is associated with PSR\,B1823-13 \citep{xmm_1823}.)
\emph{Since most massive star formation is taking place in inhomogeneous molecular clouds,
we expect to see several offset PWN, which can be studied with the high resolution imaging capabilities of {\it HESS},
since the {\it HESS} emitting electron lifetimes can be longer than the epoch of PWN shift.}

A review of the {\it HESS} telescopes and basic overview of results are given by G. Hermann in these proceedings.
In this paper we concentrate on the PWN results of {\it HESS} and what we learn from these results,
combined with multiwavelength information. At the time of the writing of this paper, a number of PWN has been
detected by {\it HESS}, but we only discuss the PWN of
SNR\,G\,0.9+0.1, PSR\,B1509-58, and PSR\,B\,1823-13.

\section{HESS\,J\,1747-281 = PWN of SNR\,G\,0.9+0.1}
Unresolved TeV emission from SNR\,G\,0.9+0.1 was detected after 50 hours of live-time observations \citep{g09}. The
image in Figure 1a shows that the TeV source is coincident with the PWN, and is
smaller than the shell size associated with this composite SNR. This source is
one of the weakest TeV sources detected so far, with an integral flux above 200 GeV of
$F(>200\,{\rm GeV})=(5.7\pm 0.7_{stat}\pm 1.2_{sys})\times 10^{-12}$ cm$^{-2}$s$^{-1}$
and a photon spectral index of $\Gamma_{HESS}=(2.40\pm 0.11_{stat}\pm 0.20_{sys})$. 
This spectrum is shown in Figure 1b \citep{g09}.

\begin{figure}[t]
  \includegraphics[height=.35\textheight]{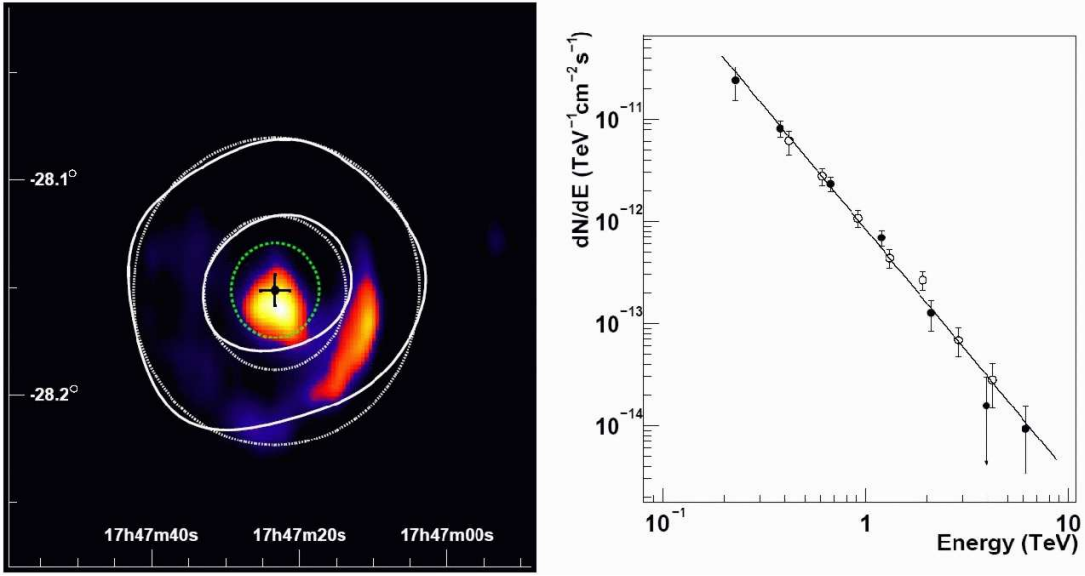}\label{fig1}
  \caption{(a) Left: The 90 cm radio flux map of G\,0.9+0.1 with HESS confidence contours: white solid (dashed) circles: observed (simulated point spread function) 40\% and 80\% peak brightness; green dashed circle: 95\% upper limit on {\it HESS} source size.   (b) Right: The {\it HESS} photon spectrum of G0.9+0.1.}
\end{figure}

de Jager et al. \citep{d95} derived an analytical expression for the TeV $\gamma$-ray spectrum in terms of the
X-ray spectrum, assuming that the X-ray spectrum with energy index $\alpha_x$ (between 0 and 2.5)
extends to the EUV/soft X-ray range, since synchrotron
emitting electrons producing such soft emission also IC scatter CMBR photons into the TeV range for
a magnetic field strength in the range $\sim 10\mu$G. The IC $\gamma$-ray energy 
spectrum (with $E_{\rm TeV}$ the $\gamma$-ray energy in units of TeV) at earth, written
in terms of the observed monochromatic X-ray flux $S_{1\,{\rm keV}}$ at 1 keV is then
(in units of TeV/cm$^2$/s/TeV)
$$
S_{2.7\rm K}(E_{\rm TeV})=6.6\times 10^{-17}\frac{(1.4\times 10^{-5})^{\alpha_x}}{B_{\perp}^{(\alpha_x+1)}}exp\left[2.2\alpha_x-0.126\alpha_x^2\right]S_{1\,{\rm keV}}
C_{2.7\rm K}E_{\rm TeV}^{-\alpha_x},
$$
with the first-order Klein-Nishina correction function given by
$C_{2.7\rm K}=[1-(2.4+3.9\alpha_x+0.42\alpha_x^2)(0.03E_{\rm TeV})^{1/2}].$
If IC emission on the 25K galactic (or swept-up) dust component 
(with density $U_{25}$ in units of eV/cm$^3$) is important, we
have to add the additional IC TeV flux given by the expression:
$$
S_{25\rm K}(E_{\rm TeV})=0.023(25/2.7)^{\alpha_x}(\alpha_x+3)(\alpha_x+4)U_{25}
\left[\frac{S_{2.7\rm K}(E_{\rm TeV})}{C_{2.7\rm K}}\right] C_{25\rm K},
$$
with the first-order Klein-Nishina correction now larger \citep{d95}
$C_{25\rm K}=[1-(2.4+3.9\alpha_x+0.42\alpha_x^2)(0.09E_{\rm TeV})^{1/2}].$
This correction function is only valid for $C<0.5$ and higher order
corrections are then necessary for larger energies.
  
Milky Way type galaxies show that the IR emission from dust dominates
the starlight component near the galactic center
(R. Tuffs \& C. Popescu, personal communication, 2005),
contrary to the assumption in \citep{g09}, where the starlight was
assumed to dominate the dust component. This observation
also makes the interpretation of the TeV spectrum simpler as discussed below:
 
{\it XMM Newton} observations of G\,0.9+0.1 showed that the energy index
$\alpha_x=\Gamma_X-1$ increases with increasing radius (possibly due to synchrotron cooling),
giving a value of $\alpha_x= 1.4\pm 0.2$ for an outer shell of radius 1.8 to 3.5 pc \citep{g09_xmm}, which
is larger compared to the 0.6 pc size of the Crab X-ray nebula. It is reasonable
to assume that the field strength in this outer region would be smaller compared to
the compact core, so that most electrons have accumulated here over the pulsar lifetime,
resulting in the observed IC TeV emission. The 95\% upper limit on the {\it HESS} source
radius (containing 68\% of the events) is also 3.2 pc \citep{g09} (assuming a
distance of 8.5 kpc to the galactic center.) The {\it HESS} source therefore
includes the total X-ray PWN and we expect most {\it HESS} emitting electrons
to be concentrated in the region where the field is lowest and volume is
largest.

It also appears to be valid to extrapolate
the X-ray spectrum into the EUV range, since the spectral break between radio and
X-rays is only observed around $10^{11}$ to $10^{12}$ Hz \citep{sid00}.
The photon index of TeV emission should therefore also be $\Gamma_{\it HESS}\sim a_x+1=2.4$
(as observed.) Assuming a monochromatic flux of $S_{1\,{\rm keV}}=0.0014$ 
cm$^2$/s/keV at 1 KeV \citep{g09_xmm}, we calculate a magnetic field 
strength in the large outer volume of the PWN as 10$\mu$G, assuming a 25K photon density 
of 1 eV/cm$^3$ in this region of the galaxy. However, a 2.3 times higher density would give a
field strength of $\sim 14\mu$G. 

\emph{Thus, the agreement between the X-ray and TeV spectral
indices (within uncertainties) exclude the need to invoke a dominant central starlight
component, consistent with observations of other Milky Way type galaxies.} Furthermore, 
a more detailed MHD approach is needed to put constraints on the injection
parameters of the unseen pulsar.


\section{HESS\,J1514-591 = PWN of PSR\,B1509-58}
This source was easily detected by {\it HESS}, with a significance of $25\sigma$ after
22 hours of observation \citep{hess_1509}. Good statisics allowed the measurement of the TeV spectrum
with high precision: The monochromatic flux at 1 TeV is
$(5.7\pm 0.2_{stat}\pm 1.4_{sys})\times 10^{-12}$ cm$^{-2}$s$^{-1}$TeV$^{-1}$, and the
power law spectral index is $\Gamma_{HESS}=(2.27\pm 0.03_{stat}\pm 0.20_{sys})$
for the range 0.28 to 40 TeV. The power law is unbroken \citep{hess_1509}.

\begin{figure}[t]
  \includegraphics[height=.3\textheight]{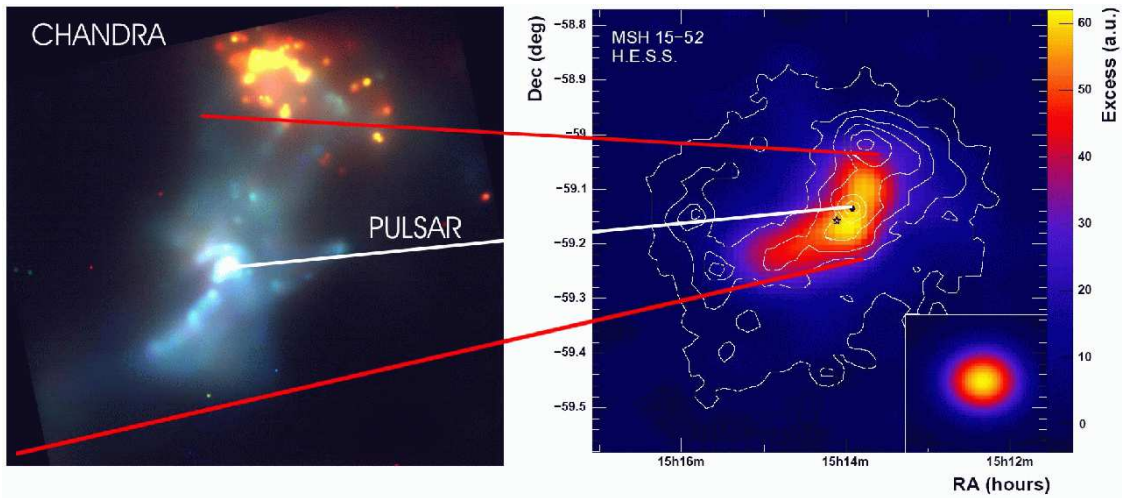}\label{fig2}
  \caption{(a) Left: The {\it Chandra} image of the PWN of PSR\,B1509-58, indicated 
  by the extended blue jet and compact torus perpendicular to the jet, centered on the pulsar. 
  The red structure at the top is the thermal nebula RCW\,89, undetected by {\it HESS}
  \citep{chandra_1509}. 
  (b) Right: The {\it HESS} image of the PWN of PSR\,B1509-58, showing a similar 
  jet structure as indicated by the projection from {\it Chandra} 
  (red and white lines) and the {\it ROSAT} PSPC contour overlay (white contours) \citep{hess_1509}.
  }
\end{figure}

From Figure 2 we see that the TeV morphology follows the X-ray morphology
in terms of the jet-like structure: 
This pulsar also shows a torus and jet, with the torus clearly resolved by {\it Chandra}
\citep{chandra_1509}.
Contrary to other compact PWN, the jet dominates the toroidal component, and the
same is seen in TeV $\gamma$-rays.  
\emph{This is the first time that an astrophysical jet has been resolved in the 
$\gamma$-ray domain.} The TeV emission extends to a radius of $0.3^\circ$ 
(or 25 pc) from the pulsar, assuming a distance of 5 kpc to the pulsar

Whereas the jet close to the pulsar reveals a photon index of 
$\Gamma_{\rm cn}=1.6\pm 0.1$
(with "cn"="compact nebula" derived from feature "C" of \citep{chandra_1509}), 
the diffuse emission up to 0.3 degrees reveal a spectral index of 
$\Gamma_{\rm en}=2.04\pm 0.01$ ("en"="extended nebula", \citep{sax_1509}). This
steepening (by $\Gamma_{\rm en}-\Gamma_{\rm cn}= 0.44\pm 0.10$) is consistent 
with synchrotron cooling.
The TeV emission corresponds to this extended cooled region and the
additional steepening $\Gamma_{\rm HESS}-\Gamma_{\rm en}=2.27-2.04=0.23$
can be explained by the addition of a 25K dust component by
galactic and swept-up dust in the SNR of this pulsar \citep{duplessis,hess_1509}.
A discussion of the cooling break will be given below. 

Aharonian et at. \citep{hess_1509} found that galactic dust emission at a level of 
2.3 eV/cm$^3$ could explain the observed {\it HESS} spectrum, assuming a
mean magnetic field strength of 17$\mu$G in this extended region.
Before the {\it HESS} era, Gaensler et al. \citep{chandra_1509} 
argued that the spectral break due to synchrotron
losses should be just below 1 keV (for an $8\mu$G field), whereas the
field strength of 17$\mu$G derived from TeV observations predicts a spectral break
near 3 TeV, which is not observed \citep{hess_1509}.
Note however that this cooling break scales as $B^{-4}T^{-2}$ in
the Thomson limit, and that the nebular field strength was larger during earlier epochs
(i.e. expansion causes the field strength to decrease with time).
Just a factor two larger $B$ (without changing the spindown age),
during earlier epochs, will already push the spectral break down
by a factor of 16. The result is then an unbroken power law down to
at least 0.3 TeV, as observed. \emph{Thus, more realistic PWN models 
should allow for field evolution with time, whereas constant
field models may predict a TeV spectral break, which is not observed.}

\section{HESS\,J1825-137 =(?) G\,18.0-0.7 = PWN of PSR\,B1823-13}
The source HESS\,J1825-137 was discovered as part of the {\it HESS} galactic
plane survey \citep{hess_plane}. It is resolved with an RMS radius of $\sim 10'$, and
the 101 ms period Vela-like pulsar, PSR\,B1823-13, is the only candidate source
within the area resolved by {\it HESS}.
The {\it HESS} Collaboration recently made a case for the association of HESS\,J1825-137
with G\,18.0-0.7, based on morphological and spectral considerations \citep{hess_1825}.

Despite the fact that the X-ray PWN G\,18.0-0.9  is located
south of this 21 kyr old pulsar, the association 
between the pulsar and the X-ray PWN is considered to be firm \citep{xmm_1823}.
Figure 3a shows an excess slice plot taken along the north-south direction,
showing this assymetry. Gaensler et al. \citep{xmm_1823} invoked the
Vela X type explanation of Blondin et al. \citep{b01} to explain the 
southwards offset of the PWN relative to PSR\,B1823-13.
From Figure 3a we can see that the bright X-ray compact nebula 
(containing freshly injected, uncooled electrons) is symmetric with respect
to the pulsar, whereas the older (cooled) electrons result in a southward offset
X-ray nebula. The TeV nebula is also shifted to the south (Figure 3b), with
an extension ($\sim 0.5^\circ$), which is even larger compared to X-rays. This
is expected based on the longer synchrotron lifetimes of TeV vs X-ray emitting electrons.

The $0.5^\circ$ diameter of the TeV source translates to a
radius of $R_{\mathrm PWN}\sim 17d_4$ pc for a distance of $d=4$ kpc. 
The SNR forward shock radius is then expected to be  
$R_{\mathrm SNR}\sim 4R_{\mathrm PWN}=70d_4$ pc \citep{vds01}, which can be reached if we
invoke a relatively large initial spindown power and pulsar braking index less than
the canonical value of $n=3$, to increase the age of the SNR.
The Sedov-Taylor SNR size would also increase with increasing age, giving a radius of 
$$R_{\mathrm SNR}=78{\mathrm pc}~
\left(\frac{E_{\mathrm SN}}{10^{51}{\mathrm ergs}}\frac{0.003\,{\mathrm cm^{-3}}}{N}\right)^{0.2}
\left(\frac{1}{n-1}\right)^{0.4}.$$
Here $E_{\mathrm SN}\sim 10^{51}$ ergs is the SN explosion energy and $N\sim 0.003$ cm$^{-3}$
is the density of the hot phase of the ISM. A pulsar braking index
of $n=2$ (``LMC pulsar'' type) will then increase the age of PSR\,B1823--13 to $\sim 42$ kyr,
which would imply a size consistent with the predicted size of the unseen SNR shell of $\sim 70$ pc.


\begin{figure}[t]
  \includegraphics[height=.25\textheight]{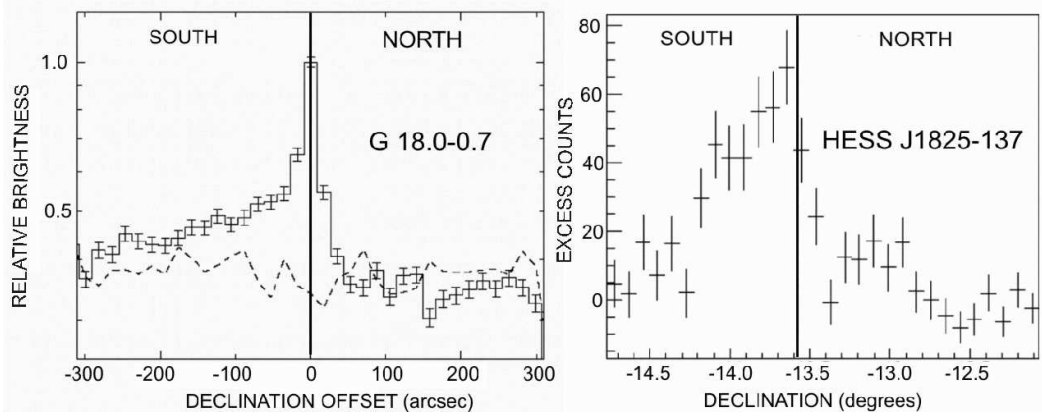}\label{fig3}
  \caption{(a) Left: The {\it XMM} north-south excess slice of the PWN of PSR\,B1823-13
  (from \citep{xmm_1823}), 
  (b) Right: similar north-south slice of HESS\,J1825-137, also showing emission
  south of the pulsar, but over a much larger size (from \citep{hess_1825}.) 
  }
\end{figure}

\bibliographystyle{aipproc}
\bibliography{Master}

\end{document}